\documentstyle[aps,epsfig]{revtex}

\begin{document}

\title{Bose-Einstein Condensation on inhomogeneous complex networks} 
\author{R. Burioni$^{1,2}$  \footnote{burioni@fis.unipr.it}, 
D. Cassi$^{1,2}$ \footnote{cassi@fis.unipr.it}, M. Rasetti $^{1,4}$ 
\footnote{rasetti@polito.it}, P. Sodano $^{1,3}$ 
\footnote{Pasquale.Sodano@pg.infn.it} and A. Vezzani$^{1,2}$ 
\footnote{vezzani@fis.unipr.it}} 
 
\address{ $^1$ Istituto Nazionale Fisica della Materia (INFM)\\ 
$^2$ Dipartimento di Fisica, Universit\`a di Parma, parco Area delle Scienze 7A 
43100 Parma Italy\\ $^3$ Dipartimento di Fisica, Universit\`a di Perugia, via A. 
Pascoli 06123 Perugia Italy\\ $^4$ Dipartimento di Fisica, Politecnico di 
Torino, corso Duca degli Abruzzi 24 10129 Torino Italy} 
\maketitle 
 
\begin{abstract} 
The thermodynamic properties of non interacting bosons on 
a complex network can be strongly affected by topological 
inhomogeneities. The latter give rise to anomalies in the  
density of states that can induce Bose-Einstein condensation in  
low dimensional systems also in absence of external confining potentials. 
The anomalies consist in energy regions composed of an 
infinite number of states with vanishing weight in the thermodynamic 
limit. We present a rigorous result providing the general conditions for the 
occurrence of Bose-Einstein condensation on complex networks in 
presence of anomalous spectral regions in the density of states. 
We present results on spectral 
properties for a wide class of graphs where the theorem applies.  
We study in detail an explicit geometrical realization, the comb lattice, 
which embodies all the relevant features of this effect and which can be 
experimentally implemented as an array of Josephson Junctions.  
 
\end{abstract} 
 
\section{Introduction} 
 
The impressive experimental demonstration of Bose-Einstein condensation  
(BEC) has stimulated a wealth of theoretical work aimed  
at a better understanding of its basic mechanism and of its potential 
consequences for the engineering of quantum devices \cite{1}.   
 
A well known general result \cite{2} is that for an ideal gas of Bose  
particle, BEC does not occur for homogeneous systems in dimension  
$d \leq 2$; the same is true for non interacting bosons on regular  
periodic lattices. The result cannot be extended to more  
general discrete structures lacking translational invariance, which, 
in principle, can now be implemented 
by combining quantum devices in non conventional discrete geometrical  
settings.   
 
The first important problem one is faced with is the characterization  
of the geometrical properties of a general network, i. e. a graph.  
On one hand, one is interested in describing the complex geometry of a sample 
and identifying its effects on physics. On the other, once the effects 
of geometry are known, one can use the geometrical 
setting to tune the physical properties, by a sort of engineering 
in network design. As we will see, this can lead to effects 
analogous to the introduction of an external confining potential 
for regular geometries. 
 
The problem of describing the large scale geometry of graphs by 
an effective parameter that generalize the Euclidean dimension of regular 
lattices has been successfully solved by the introduction of the 
spectral dimension $\bar{d}$ \cite{ao,univ}, which can be experimentally 
measured \cite{misur} and rigorously defined by graph theory \cite{hhw}. 
However, one expects the influence of topology to be richer and 
more complex on discrete structures, due to the possible relevance of local 
geometrical details, in addition to the large scale structure described by 
dimensionality. 
 
In this direction, recent works on BEC on inhomogeneous networks  
\cite{beccomb} have put into evidence that strong inhomogeneities 
can give rise to condensation at finite temperature for a system of non 
interacting bosons in absence of an external confining potential even in a 
low-dimensional structure as a comb lattice, where $\bar d = 1$. 
This phenomenon arises from a peculiar part of the density of states  
on inhomogeneous networks we shall refer to as {\it hidden spectrum} 
\cite{hidden}. This hidden spectrum 
consists  of an energy region filled by a finite or infinite number of 
states which do not 
contribute to the normalized density in the thermodynamic limit. 
Hidden spectra do not usually affect bulk thermodynamic quantities but, 
as we shall show in the following, can have dramatic effects since bosonic 
statistics allows the macroscopic occupation of a single quantum state. 
 
In this paper we shall determine the conditions under which,  
even for low dimensional systems, non-interacting bosons on a  
general discrete network  may lead  
to BEC \cite{hidden}. We shall see that this is indeed possible  
if one resorts to a suitable discrete inhomogeneous ambient space on which  
bosons are defined;  
this is very desirable in view of the engineering of quantum 
devices. 
 
Bosonic models over generic discrete structures can be made 
experimentally accessible through the realization of arrays of  
Josephson junctions (JJA) \cite{ba,si}. The latter are devices  
that can be engineered to realize a variety of non-homogeneous  
patterns. We shall see indeed that classical \cite{ba,si} JJA  
arranged in a non-homogeneous network provide a relevant example  
of the proposed mechanism for BEC, leading to a spatial condensation  
of bosons in a single state.   
 
The model Hamiltonian describing the low-temperature JJA is  
the Bose-Hubbard (BH) model \cite{bh}.  
On a generic graph, the Hamiltonian is given by 
\begin{equation} 
H^{BH}=U \sum\limits_i n_i^2 + \sum\limits_{ij} A_{ij}  
\Big( V n_i n_j - J(a_i^{\dag} a_j+ a_j^{\dag} a_i) \Big) 
\end{equation}
where $A_{ij}$ is the adjacency matrix of the graph:  
$A_{ij}=1$ if the sites $i$ and $j$  
are nearest neighbours and $A_{ij}=0$ otherwise;  
$a_i^{\dag}$ creates a boson at site $i$ and  
$n_i \equiv a_i^{\dag} a_i$. The phase diagram reflects the  
competition between the boson kinetic energy (hopping term)  
and the repulsive Coulomb interaction. In a realistic experimental  
setup \cite{exp}, the parameters $U$ and $V$ depend  
on the ratio between the intergrain capacitance $C$ and the  
gate capacitance $C_0$, while the parameter $J$ describes  
Cooper pair hopping. Classical JJA are obtained  
when $U,V << J$ (i.e. $C/C_0 \to 0$); in this limit  
the hopping term dominates the physics of the system.  
 
Recently, the BH model has been shown to describe the dynamics of  
ultracold bosonic atoms on a particular regular discrete geometry,  
the optical lattice \cite{zoller}. In this context  
the BH model describes the hopping of bosonic atoms between low  
vibrational states of the optical lattice; the parameters  
$U$ and $J$ depend on the power of the laser beams used to realize   
the lattice. The dynamics of a BEC in such a kind of periodic potential  
can thus be described by a discrete nonlinear Schr\"odinger equation  
\cite{discrete}. Also for atomic Bose systems,  
the experiments have reached  
an high level of accuracy with easily tunable external parameters  
and accurately tailored trapping profiles \cite{anderson}.  
 
In this paper we will consider a model of pure hopping of non 
interacting bosons on general non regular geometries, such as  
complex networks. In particular,  
we will analyse the possibility of having a 
Bose-Einstein condensate at finite temperature, providing a general  
condition for BEC on a general network \cite{hidden}.  
We will describe in detail some geometrical 
settings where inhomogeneities act as a confining 
potential, giving rise to BEC even in low dimensions. 
 
The extension of this analysis to the full BH model on complex networks  
is not trivial and it is still the object of on-going investigations. 
 
The paper is organized as follows: in section II we introduce the 
mathematical notations and some basic concepts in algebraic graph theory; 
in section III we define the pure hopping model and prove a general 
theorem giving the most general condition for the existence of BEC at 
finite temperature for non interacting bosons on a complex network. 
We discuss in detail the presence of ``hidden'' regions in the density of 
states in the thermodynamic limit; 
in section IV 
we apply the theorem to the case of the comb graph showing 
that BEC at finite temperature is present in $1$-dimensional structures as a 
consequence of the effects of the inhomogeneous geometry on the density of  
states; finally in section V 
we consider a wide class of networks where the relation between 
BEC and complex geometry can be analyzed in detail. 
 
\section{Complex networks and graphs} 
 
On complex networks it is not possible to introduce 
important mathematical tools typical of Euclidean lattices,  
such as the Fourier transform and the reciprocal lattice.  
Nevertheless general and effective 
mathematical techniques are provided by graph theory and the algebraic approach 
to graph topology \cite{alg}. In this section we briefly recall some basic 
definitions and notations. 
 
A graph $G$ is a countable set $V$ of vertices $i$ connected pairwise by a set 
$E$ of unoriented links $(i,j)=(j,i)$. In general sites represent degrees of 
freedom and links represent interactions between them. In particular, for  
bosonic models, vertices can be regarded as Josephson junctions or as sites in  
the optical networks, while edges represent the hopping probabilities  
between sites. 
  
A subgraph $S$ of $G$ is a graph whose set of vertices $V' \subseteq V$ 
and whose set of links $E'\subseteq E$.  
A path in $G$ is a sequence of consecutive links $\{(i,k)(k,h)\dots (n,m)(m,j)  
\}$ and a graph is said to be connected, if for any two points $i,j \in V$  
there is always a path joining them. In the following we will consider connected  
graphs. Every connected graph $G$ is endowed with an intrinsic metric generated  
by the chemical distance $r_{i,j}$, which is defined as the number of links in  
the shortest path connecting vertices $i$ and $j$. On a graph  
the generalized Van Hove sphere $S_{o,r}\subset G$ of center $o$ and radius $r$  
is the subgraph of $G$ containing all $i\in G$ whose distance from $o$ is $\leq 
r$ and all the links of $G$ joining them. We will call $N_{o,r}$ the number of  
vertices contained in $S_{o,r}$.  
The Van Hove spheres play an important role in the study of the  
thermodynamic limit on infinite graphs. The physical quantities are usually evaluated  
by first considering the models restricted to the finite spheres  and then  
taking the thermodynamic limit by letting $r\to \infty$. The average in the thermodynamic  
limit $\overline{\phi}$ of a bounded function defined on the vertices $\phi(i)$ is:  
\begin{equation}  
\overline{\phi}\equiv \lim_{r\rightarrow\infty}  
{\displaystyle \sum_{i\in S_{o,r}} \phi(i) \over \displaystyle N_{o,r}}~~.  
\label{deftd}  
\end{equation} 
With this definition, the measure $|V'|$ of a subset $V'$ of $V$ is the average value of its  
characteristic function $\chi_{V'}(i)$, defined by $\chi_{V'}(i)=1$ if $i\in V'$ 
and $\chi_{V'}(i)=0$ otherwise. It can be shown \cite{rimtim} that for graphs  
with polynomial growth, i.e. when $N_{r,o}\sim r^p$ for $r\to \infty$, the  
thermodynamic limit is independent from the choice of the center of the sphere 
$o$. Since all realistic networks which can be embedded in a  
3-dimensional space have polynomial growth, we restrict to this class  
graphs and we drop the index $o$ of the  
center of the sphere. The graph topology is algebraically described by its  
adjacency matrix:  
\begin{equation} 
A_{ij}=\left\{  
\begin{array}{cl}  
1 & {\rm if } \ (i,j) \in E \cr  
0 & {\rm if } \ (i,j) \not\in E \cr  
\end{array}  
\right .  
\label{defA}  
\end{equation}  
  
On disordered structures, a useful generalization of the adjacency matrix can be considered by  
introducing a bounded distribution of hopping probabilities $t_{ij}$ which  
differs from link to link: 
\begin{equation}  
t_{ij}=t_{ji}\left\{ 
\begin{array}{cl}  
\not= 0 & {\rm if } \ A_{ij}=1 \cr  
=0 & {\rm if } \ A_{ij}=0 \cr  
\end{array}  
\right .  
\label{deft}  
\end{equation}  
with $\sup_{(i,j)}|t_{ij}|<\infty$.  
 
The Laplacian matrix $L_{ij}$ is defined by:  
\begin{equation}  
L_{ij} = z_i \delta_{ij} - A_{ij}  
\label{defDelta}  
\end{equation}  
where  $z_i=\sum_j A_{ij}$, the number of nearest neighbours of $i$, is called  
the coordination number. We will consider graph with bounded coordination  
number, i.e. with $\max_i z_i<\infty$.  
  
$L_{ij}$ is the generalization to graphs of the usual Laplacian on a lattice,  
where $z_i=z$, $\forall i$. $L_{ij}$ has some important spectral properties: its  
spectrum is real, non-negative and bounded. In particular $0$ is a simple 
eigenvalue of $L_{ij}$ corresponding to the constant eigenvector. Notice  
that, while on a regular lattice $L_{ij}$ is diagonalized by the Fourier 
transform, this is not the case for a generic graph.  
  
A fundamental problem in graph theory is the characterization  
of the large scale geometry of the graph by the  
definition of a dimension which generalizes to infinite graphs the  
usual Euclidean dimension $d$ of lattices. The spectrum of the Laplacian operator  
defines the spectral dimension $\bar{d}$,  
which has been shown to be the extension of $d$ in many physical problems.  
On a infinite graph, the 
spectral density of the Laplacian matrix is denoted by $\rho(l)$. The spectral dimension  
$\bar{d}$ is defined by the asymptotic behaviour of  
$\rho(l)$ at low eigenvalues \cite{ao,univ}:  
\begin{equation}  
\rho(l)\sim l^{\bar{d}/2-1}\ \ \ \ \ {\rm for\ \ } \l \to 0^+  
\label{defdt}  
\end{equation}  
Interestingly, $\bar{d}$ can be a real number, it can be experimentally  
measured \cite{misur}, it depends only on the large scale topology of the graph  
\cite{univ} and it is the direct extension to graphs of the Euclidean dimension  
in many physical problems. In particular, the long time  
asymptotic diffusion of classical particles \cite{ao}, the low frequencies 
spectrum of classical harmonic oscillators \cite{ao}, the zero mass limit of the  
Gaussian model \cite{univ,hhw} and the existence of spontaneous magnetization in 
continuous symmetry spin models \cite{gfss} are determined by the spectral  
dimension.

\section{BEC on complex networks: the general theorem}  
  
The Hamiltonian for non interacting bosons on a graph ${G}$ is:  
\begin{equation}  
{H}=\sum_{i,j\in V} h_{ij}a^{\dag}_ia_j 
\label{ham}  
\end{equation}  
where $a^{\dag}_i$ and $a_i$ are the creation and annihilation operator  
at site $i$, with $[a_i,a^{\dag}_j]=\delta_{ij}$.  
The Hamiltonian matrix $h_{ij}$ is defined by:  
\begin{equation}  
h_{ij}= t_{ij} + \delta_{ij} V_i  
\end{equation}  
The term $t_{ij}$, defined in (\ref{deft}), describes hopping between nearest  
neighbours sites. The diagonal term  
$V_i$ takes into account a potential at site $i$ and it satisfies  
a boundedness condition: $\sup |V_i|< \infty$. In particular, the theorem 
holds also when $V_i=0$, i.e. when there is no on site potential. 
  
The thermodynamic limit on the graph is studied by first 
considering the finite Van Hove sphere $S_r$ and then letting $r  
\to \infty$. This definition of the thermodynamic limit is very general and  
it applies to a general discrete structure. On graphs with symmetries  
(lattices, fractals, bundled structures, etc.) different shapes of Van Hove  
spheres can be introduced, preserving the symmetry of the network. It can be  
shown that for regular choices of the spheres the properties of the model for $r  
\to \infty$ do not depend on the shape of the sphere.  
The Hamiltonian (\ref{ham}) restricted to the sphere is:  
\begin{equation} 
{{H}}^{S_{r}}=\sum_{i,j\in S_r} h^{S_{r}}_{ij}a^{\dag}_ia_j  
\label{hamsr}  
\end{equation}  
where $h^{S_{r}}_{ij}=h_{ij}$ if $i$ and $j$ belong to $S_r$ and  
$h^{S_{r}}_{ij}=0$ otherwise.  
  
For each finite sphere of radius $r$ let us consider the eigenvalues  
equation:  
\begin{equation}  
\sum_{j\in S_r} h^{r}_{ij}\psi(j)=E \psi(i)  
\label{eigen}  
\end{equation} 
  
The normalized density of states $\rho^r(E)$ on $S_r$ is defined as: 
\begin{equation}  
\rho^r(E)={1\over N^r}\sum_{k} \delta(E-E^r_k)  
\label{densityr}  
\end{equation}  
where $E^r_k$ are the $N^r$ eigenvalues of $H^r_{ij}$.  
  
We define $\rho(E)$ to be the density of states of $h_{ij}$  
in the thermodynamic limit if:  
\begin{equation} 
\lim_{r\to\infty} \int |\rho^r(E)-\rho(E)| dE =0  
\label{dens}  
\end{equation}  
Let $E_m \equiv {\rm Inf} ({\rm Supp}(\rho(E)))$, where  
${\rm Supp}(\rho(E))$ is the support of the distribution $\rho(E)$.  
On a graph, the asymptotic behavior at low energies of the density of states is  
described by an exponent $\alpha$:  
\begin{equation}  
\rho(E) \sim (E-E_m)^{{\alpha \over 2}-1} \ \ \ \ \ \ {\rm for} \ E\to E_m  
\label{defd} \end{equation}  
The exponent $\alpha$ in general depends both on graph topology and on  
the external potential. Interestingly, when $h_{ij}=L_{ij}$ and when the external potential 
$V_i$ satisfy suitable regularity conditions, $\alpha$ coincides with the spectral  
dimension $\bar{d}$. In particular this happens when $V_i=0$. 
A {\it hidden} region of the spectrum is an energy  
interval $[E_1,E_2]$ such that  
$[E_1,E_2]\cap {\rm Supp}(\rho(E))= \emptyset$ and  
$\lim_{r\to \infty} N^r_{[E_1,E_2]}>0$, where  $N^r_{[E_1,E_2]}$ is the number  
of eigenvalues of $H^r_{ij}$ in the interval $[E_1,E_2]$. Notice that in general  
$N^r_{[E_1,E_2]}$ can diverge for $r\to \infty$ and the eigenvalues can become  
dense in $[E_1,E_2]$ in the thermodynamic limit. Therefore this condition not  
only includes the trivial case of discrete spectrum but is far more general; an  
interesting example of this behaviour is found in the comb 
lattice without external potential \cite{beccomb} which will be studied in detail  
in the next section.  
  
We now define the lowest energy level for the sequence of  
densities $\rho_r(E)$, setting  
$E_0^r={\rm Inf}_k (E_k)$ and  
$E_0=\lim_{r \to \infty} E_0^r$. In general, $E_0\leq E_m$.  
If $E_0<E_m$, then $[E_0,E_m]$ is a hidden region of the spectrum,  
which will be called hidden low-energy spectrum.  
  
In the following we will consider models at fixed filling  
$f=N/N_r$ ($N$ is the number of particles in the system). In the grand canonical  
ensemble the equation that determines the fugacity $z$ in the thermodynamic 
limit is:  
\begin{equation}  
f= \lim_{r\to\infty} 
\int {\rho^r(E) dE\over z^{-1}e^{\beta E}-1}  
\label{eq1}  
\end{equation}  
Setting $E_0=0$ we have that $0\leq z \leq 1$.  
  
The integral in the equation (\ref{eq1}) can be divided into two sums, the  
first one considering the energies smaller than an arbitrary constant  
$\epsilon$ and the second the energies larger than $\epsilon$:  
\begin{equation} 
\int {\rho^r(E) dE\over z^{-1}e^{\beta E}-1}=  
\sum_{k=0}^{E_k\leq \epsilon}  
{1 \over z^{-1}e^{\beta E_k^r}-1}+  
\int _{E>\epsilon} {\rho^r(E) dE\over z^{-1}e^{\beta E}-1}  
\label{eq2}  
\end{equation}  
  
We define:  
\begin{equation}  
n_{\epsilon}^r \equiv  
\sum_{k=0}^{E_k\leq \epsilon}  
{1 \over z^{-1}e^{\beta E_k^r}-1} 
\label{ne}  
\end{equation} 
as the fraction of particles with energy smaller than  
$\epsilon$. Bose-Einstein condensation occurs in this systems if it exists a  
critical temperature $T_c>0$ such that, for any $T<T_c$,  
$n_{\epsilon} \equiv \lim_{r\to\infty}n_{\epsilon}^r>k>0$, for all  
$\epsilon>0$; i.e. $n_0\equiv\lim_{\epsilon\to 0}n_{\epsilon}=k>0$.  
  
From definition (\ref{ne}), $n_0$ can be strictly positive  
only if $\lim_{r\to\infty}z(r)=1$. Indeed, if this limit is  
smaller than $1$ it follows that 
$n_{\epsilon}\leq(1-z)^{-1}\lim_{r\to\infty} N_{\epsilon}^r / N^r$,  
where $N_{\epsilon}^r$ is the number of state with energy smaller than  
$\epsilon$. If the ground state is not infinitely degenerate,  
$\lim_{\epsilon \to 0}\lim_{r\to \infty}(N_{\epsilon}^r/N^r)=0$  
and then $n_0=0$.  
  
Taking first the limit $r\to\infty$ and then  
$\epsilon \to 0$ in equation (\ref{eq2}) we obtain:  
\begin{equation}  
f  =   n_0+ \lim_{\epsilon\to 0}  \lim_{r\to\infty}  
\int_{E>\epsilon} {(\rho^r(E)-\rho(E))dE\over z^{-1}e^{\beta E}-1} 
+ \lim_{\epsilon \to 0} \lim_{r\to\infty} 
\int_{E>\epsilon} {\rho(E) dE\over z^{-1}e^{\beta E}-1}  
\label{eq4} 
\end{equation}  
Now, from the boundedness of $(z^{-1}e^{\beta E}-1)^{-1}$ for $E>\epsilon$  
and from  
the definition (\ref{dens}), the first of the two limits in the right hand side  
of (\ref{eq4}) vanishes:  
\begin{equation}  
f= n_0+ \lim_{\epsilon \to 0}  
\int_{E>\epsilon} {\rho(E) dE\over z^{-1}e^{\beta E}-1}=  
n_0+\int {\rho(E) dE\over z^{-1}e^{\beta E}-1} 
\label{eq7}  
\end{equation}  
where, again, $n_0$ can be different from $0$ only if $z=1$.  
  
The integral in equation (\ref{eq7}) is an increasing continuous  
function of $z$ with $0\leq z < 1$. If the limit:  
\begin{equation}  
f_c(\beta)=  
\lim_{z \to 1}  
\int{\rho(E) dE\over z^{-1}e^{\beta E}-1}  
\label{eq8}  
\end{equation} 
is equal to $\infty$, then $z<1$ and $n_0=0$. If the limit is  
finite, $f_c(\beta)$ is a decreasing function of $\beta$ with 
$\lim_{\beta \to \infty}f_c(\beta)=0$ and  
$\lim_{\beta \to 0}f_c(\beta)=\infty$.  
Then, for a suitable $\beta_c$, $f_c(\beta_c)=f$. For $\beta>\beta_c$,  
(i.e. $T<T_c$) $z=1$ and  $n_0=f-f_c(\beta)>0$ while for $\beta<\beta_c$,  
(i.e. $T>T_c$) $z<1$ and $n_0=0$.  
  
From the divergence or finiteness of the limit (\ref{eq8}) one obtains  
the most general condition  
for the occurrence of Bose-Einstein condensation on a graph. 
  
First, if $0=E_0<E_m$ (i.e. the system presents a low-energy hidden spectrum)  
the limit (\ref{eq8}) is finite  
and there is Bose-Einstein condensation at finite temperature:  
\begin{equation}  
f_c(\beta)\leq  
{1\over e^{\beta E_m}-1}  
\int{\rho(E) dE}={1\over e^{\beta E_m}-1}.  
\label{eq9}  
\end{equation}  
  
On the other hand, when $E_0=E_m$ the value of the limit (\ref{eq8}) 
is determined by the exponent $\alpha$. Indeed if $\alpha>2$:  
\begin{equation} 
f_c(\beta)  \leq  
\int_0^{\delta}{c_1 E^{{\alpha \over 2}-1} dE\over \beta E}+  
\int_{E>\delta} {\rho(E) dE\over e^{\beta E}-1} < \infty  
\label{eq10}  
\end{equation}  
where $\delta$ and $c_1$ are suitable constants. Therefore in  
this case Bose-Einstein condensation occurs at finite temperature.  
  
For $\alpha<2$ we have: 
\begin{equation}  
f_c(\beta) \geq  \lim_{z \to 1}  
\int_0^{\delta} { z c_1 E^{{\alpha \over 2} -1 } dE\over \beta E +1 -z}=\infty  
\label{eq11}  
\end{equation}  
and there is no Bose-Einstein condensation. When $\alpha=2$ we have to consider  
the logarithmic correction to (\ref{defd}) and it is possible to show that the  
limit (\ref{eq8}) diverges.

The Hamiltonian (\ref{ham}) describes different models of non interacting  
bosons on graphs. The simplest example is the discretization of the usual 
Schr\"odinger equation. In this case the Hamiltonian is: $h_{ij}= 
{\hbar^2 \over 2m} L_{ij}$, where $L_{ij}$ is the Laplacian operator. This is 
a pure topological model, completely defined by the graph geometry and it can be  
shown that $E_0=E_m$, i.e. there are no hidden states in the low energy  
region and $\alpha=\bar{d}$. Therefore the occurrence of BEC is determined by  
the spectral dimension (\ref{defdt}), which has been calculated for a wide class of  
discrete networks \cite{hhw,rammal,bundle}.  
  
On real condensed matter structures an interesting case is a pure hopping  
of non interacting bosons on graphs. This has been  
considered in \cite{beccomb} for the description of the Josephson junction 
arrays in the weak coupling limit with Hamiltonian matrix given  
by: $h_{ij}=-t A_{ij}$. In this case the behaviour of the model is much more  
complex and the presence of hidden spectra can give rise to Bose-Einstein  
condensation also in low dimensional structures \cite{beccomb}.  
  
Finally, the most general application of our result concerns particles  
interacting with an external potential, described by the Hamiltonian:  
$h_{ij}=-t A_{ij}+V_i \delta_{ij}$. Indeed, the introduction of a  
local potential strongly enhances the possibility of modifying the energy  
spectra in order to induce BEC in non regular geometries.  
  
In the next section, we will study in detail the pure hopping model defined on 
the comb graph, where the inhomogeneous topology produces low energy hidden  
states which give rise to Bose-Einstein 
condensation in a $1$-dimensional structure ($\alpha=\bar{d}=1$).

\section{Pure hopping models and BEC on the comb graph}  
  
The comb lattice is an infinite graph, which can be obtained connecting to each  
site of a linear chain, called backbone, a 1-dimensional chain called  
finger (see figure 1). The sites of the comb can be naturally labelled  
introducing two integer indices $(x,y)$ with $x,y\in Z$, where $x$ labels the 
different fingers and $y$ represents the distance from the backbone.  
  
A suitable definition of the finite Van Hove sphere for the comb lattice is the  
square periodic comb of $L \times L$ sites. In this case the backbone and each  
finger are rings of $L$ sites. Each vertex can be labelled by $(x,y)$,  
$x,y\in [0,1,\dots,L-1,L]$, with the conditions $(0,0)\equiv (L,0)$ and  
$(x,0)\equiv(x,L)$ in order to guarantee periodic boundaries.  
  
In the thermodynamic limit the spectrum of the infinite comb is evaluated as  
the limit for $L \to \infty$ of the spectrum of the hopping model on the finite  
$L\times L$ combs. The eigenvalue equation (\ref{eigen}) on the finite comb  
with $h_{ij}=-tA_{ij}$ with the previous labeling of sites reads: 
\begin{equation}  
-t\sum_{x',y'=0}^{L-1}\left[(\delta_{x,x'+1}+\delta_{x,x'-1})\delta_{y,0}\delta_ 
{ 0 , y ' } + (\delta_{y,y'+1}+\delta_{y,y'-1})\delta_{x,x'}\right]  
\psi(x',y')=E \psi(x,y)  
\label{eq1b}  
\end{equation}

By exploiting the translation invariance in the  
direction of the backbone, a Fourier transform in the variable  
$x$ reduces (\ref{eq1b}) to a 1-dimensional eigenvalue problem. Let us  
define: 
\begin{equation}  
\psi(k,y)=\sum_{x} e^{ikx}\psi(x,y) \label{ft1}  
\end{equation}  
with $k=2\pi n/L$, $n=1\dots L$. The eigenvalues equation (\ref{eq1b}) becomes:  
\begin{equation}  
-t\sum_{k',y'}\left[-2\cos(k)\delta_{y,0}\delta_{0,y'}\delta_{k,k'}+  
(\delta_{y,y'+1}+\delta_{y,y'-1})\delta_{k,k'}\right]\psi(k',y')=E \psi(k,y)  
\label{eq2b}  
\end{equation}  
  
Now (\ref{eq2b}) is diagonal in the variable $k$ and can be written as:  
\begin{equation} 
-t\sum_{y'}\left[-2\cos(k_0)\delta_{y,0}\delta_{0,y'}+  
(\delta_{y,y'+1}+\delta_{y,y'-1})\right] \psi(y')=E \psi(y) 
\label{eq3b}  
\end{equation}  
where $\psi(k,y)=\delta(k-k_0)\psi(y)$ with $k_0=2\pi n/L$,  
$n=0\dots L$. Equation (\ref{eq3b}) can be regarded as a 1-dimensional problem of a  
quantum particle interacting with a potential in the origin,  
$V(k_0)=-2t \cos(k_0)$. The eigenvalues and eigenvectors of (\ref{eq3b})  
are obtained by imposing the matching condition in $y=0$ for the free particle  
solutions on the chain. Indeed, for $y\not  =0$, (\ref{eq3b})  
the wave function  satisfying (\ref{eq3b}) in $y\not=0,L-1$ are: 
\begin{eqnarray}  
\psi(y)=\cos(hy+\alpha) & {\rm\ \ \ for\ \ \ } & -2t\leq E  
=-2t\cos(h)\leq 2t; \label{sol1} \\  
\psi(y)=Ae^{hy}+Be^{-hy} & {\rm \ \ \ for\ \ \ } &  
E=-t (e^h+e^{-h})<2t; \label{sol2}\\  
\psi(y)=A(-1)^ye^{hy}+B(-1)^ye^{-hy} &  
{\rm \ \ \ for\ \ \ } & E=t (e^h+e^{-h})>2t.  
\label{sol3}  
\end{eqnarray}  
Imposing $\psi(y)$ to be a solution in $y=0$ and $y=N-1$ introduces restrictions  
on the parameters $h$, $\alpha$, $A$ and $B$, and on the eigenvalues $E$.  
 
Let us first consider the conditions for (\ref{sol1}), in $y=L-1$ and $y=0$:  
\begin{eqnarray} 
-t\cos[h(L-2)+\alpha]-t\cos(\alpha)=-2t\cos(h)\cos[h(L-1)+\alpha] \label{eq4b}\\  
-t\cos[h(L-1)+\alpha]-t\cos(h+\alpha)-2t\cos(k_0)\cos(\alpha)=  
-2t\cos(h)\cos[h(L-1)+\alpha] \label{eq5b}  
\end{eqnarray}  
The system has odd solutions with $\alpha=\pi/2$  
(i.e. $\psi(i)=\sin(hi)$), $h=2\pi n/L$ and  $n=1,\dots,L/2-1$ and  
even solutions, obtained by solving:  
\begin{equation}  
-\cos(k_0) \cot(hL/2)=\sin(h) 
\label{eq7b}  
\end{equation}  
Equation (\ref{eq7b}) can be solved graphically (see figure 2)  
obtaining $L/2$ solutions. In the large $L$ limit the allowed values for  
$h$ are: $h \approx \pi(2n -1)/L$ with $n=1,\dots,L/2$. Each  
values of $k_0$ corresponds to $L-1$ solutions with energy between $-2t$ and  
$2t$.  
  
Let us now consider cases (\ref{sol2},\ref{sol3}). For $E<2t$ the wave function  
in $y=L-1$ and $y=0$ satisfies the conditions:  
\begin{eqnarray}  
-t(Ae^{-h(L-2)}+Be^{h(L-2)})-t(A+B)=-t(e^h+e^{-h})(Ae^{-h(L-1)}+Be^{h(L-1)}) 
\label{eq8b}\\  
-t(Ae^{-h(L-1)}+Be^{h(L-1)})-t(Ae^{-h}+Be^{h})-2\cos(k_0)(A+B)= 
-t(e^k+e^{-k})(A+B) \label{eq9b}  
\end{eqnarray}  
leading to:  
\begin{equation}  
\cos(k_0) \coth(hL/2)=\sinh(h)  
\label{eq10b}  
\end{equation}  
Equation (\ref{eq10b}) can be solved graphically. For $\cos(k_0)>0$,  
(\ref{eq10b}) has a real solution. In the limit $L\to\infty$ this is 
$\sinh(h)=\cos(k_0)$, corresponding to energy is  
$E=-2t\sqrt{1+\cos^2(k_0)}$. If $\cos(k_0)<0$, only the solution with $E>2t$ is  
present and for $\cos(k_0)=0$ one has the constant solution $\psi(y)=1$  
with energy $E=0$.  
  
The complete spectrum of the comb is now  
obtained by considering the eigenvalues of (\ref{eq3b}) for the $L$  
values of $k_0$. For each $k_0$ there are $L-1$ eigenvalues of type (\ref{sol1})  
with energy $E=-2t\cos(h)$ and wave functions $\psi(x,y)=e^{ik_0x}\sin(hx)$ and  
$\psi(x,y)=e^{ik_0x}\cos(h|y|+\alpha)$, with $h$ and $\alpha$ satisfying  
conditions (\ref{eq4b},\ref{eq5b}). The fraction of states in this spectral  
region is $f=L(L-1)/L^2$. Notice that $f$ tends to 1 in the thermodynamic limit. 
The normalized density of states in this spectral region is:  
\begin{equation} 
{dn \over L^2}={L d(Lh/2\pi)\over L^2 }={dE\over \pi \sqrt{4t^2-E^2}}  
\label{den1}  
\end{equation}  
with $E\in[-2t,2t]$.  
Since the fraction of states with energies $E<-2t$ or $E>2t$ vanishes in the  
thermodynamic limit, (\ref{den1}) represents the normalized density  
of states $\rho(E)$ of the pure hopping model on the comb graph.  
  
However, the lowest energy state is obtained from a solution of type 
(\ref{sol2}). These solutions appear only for $\cos(k_0)>0$ and they  
satisfy relation (\ref{eq10b}). Since the energy is a decreasing function of  
$h$, the lowest energy level is obtained for $\cos(k_0)=1$. In  
this case we have $\coth(hL/2)=\sinh(h)$, and in the thermodynamic limit this  
corresponds to $\sinh(h)=1$. From (\ref{sol2}) we get $E_0=-t\sqrt{8}$. The  
lowest energy in the normalized density of states is $E_m=-2t$ and therefore  
$E_0<E_m$. Then the eigenvalue equation (\ref{eq1b}) presents a hidden spectrum  
in the thermodynamic limit, and the pure hopping boson model on the comb lattice  
exhibits BEC at low enough temperatures. From equations  
(\ref{sol2},\ref{eq8b},\ref{eq9b}) it is also possible to show that on the  
infinite graph the eigenvector corresponding to the lowest energy eigenvalue is  
$\psi(x,y)=e^{-h|y|}$. The wave function for the condensate 
is localized along the backbone and it decreases exponentially along the finger 
(see figure 3). 
 
This model does not present an energy gap between 
$E_0$ and $E_m$, contrary to the typical instance of non interacting 
bosons trapped in a harmonic well.  
Indeed for each value of $k_0$ ($\cos(k_0)>0$) there is  
a solution of (\ref{eq10b}) with a different energy in the interval  
$[E_0, E_m]$. In a finite comb of $L^2$ sites there are $L/2$ solution of  
this type and for $L\to \infty$ these solutions fill densely the interval $[E_0  
E_m]$. Normalizing the density of states to $L$, we can obtain the  
spectral density in the thermodynamic limit for $E\in[E_0 E_m]$:  
\begin{equation} 
{dn \over L}={d(Lk_0/2\pi)\over L}={|E|dE\over 2\pi  
\sqrt{8t^2-E^2}\sqrt{E^2-4t^2}} \label{den2}  
\end{equation}  
An analogous equation holds for the spectral region at high energy  
($E=[2t,\sqrt{8}t]$), where the other hidden states appear. The  
density of states can then be represented as in figure 4, where we  
used the two different normalizations for the hidden spectra and  
for $\rho(E)$.  
  
>From the knowledge of the complete spectral density of the comb, the thermodynamic 
quantities relative to BEC such as the filling of the ground state as a  
function of $T$, the specific heat and the critical temperature as a function of  
$f$ can be analytically calculated \cite{beccomb}. 
  
\section{Bose-Einstein Condensation on graphs: further examples} 
  
The spectrum of the pure hopping model can be studied in detail in a wide  
class of fractals and inhomogeneous discrete structures.  
In particular, the behaviour observed in the comb lattice is typical  
of bundle structures \cite{bundle}. This class of graphs are  
obtained by a ``fibering'' procedure, i.e. attaching a fiber graph to  
every point of a ``base'' graph. They are characterized  
by a hidden low-energy region, and from  
the general result of section III they exhibit BEC at finite temperature. 
Moreover, as in the comb graph,  
the wave function of the condensate is localized  
along the base and presents a fast decay along the fibers.  
An example of this behaviour is  
found in the brush graph, which is illustrated in figure 5.  
  
An interesting example of a graph where the geometrical setting gives  
rise to an isolated eigenstate in the low-energy region of the pure  
hopping model is the star graph. The effect, which is analogous to  
that caused by an impurity on a line, here is not produced by the presence  
of an external potential but simply by the topology of the system.  
Interestingly, the wave function is here completely localized in the center 
of the star, with an exponential decay along the arms.  
In figure 6 the star graph with the ground state and the spectrum of the pure  
hopping model are shown.  
  
Graphs with constant coordination number are typical  
examples in which the pure hopping model do not present hidden low energy  
regions. This is due to the fact that for this class of graphs  
the spectrum of the model can  
be obtained from that of the Laplacian matrix by a shift of the zero in  
the energy. 
The existence of BEC is therefore determined by the spectral dimension  
$\bar{d}=\alpha$ of the graph, describing the density of eigenvalues  
of the Laplacian at low energy. This parameter can be exactly calculated for a  
wide class of discrete structures. On lattices, $\bar{d}$  coincides  
with the usual Euclidean dimension and one recovers the classical result for  
BEC on translation invariant structures. For  
exactly decimable graphs \cite{hhw} (e.g. the Sierpinski gasket \cite{rammal}  
and the T-fractal, see figure 7)  
where one always has $\bar{d}<2$ there is no Bose-Einstein condensation.  
 
A fundamental property of the spectral dimension is its independence  
from the local details of the graph, i.e. universality \cite{univ}. The value 
of $\bar{d}$ and the behaviour of the density of eigenvalues in the low  
energy region is not changed under a wide class of transformations, called 
isospectralities, which can strongly modify the geometry of the graph. A  
simple consequence of this property is that if we consider the pure hopping  
model on a graph with constant coordination number,  
which differs from a graph of known dimension $\bar{d}$ by an isospectrality,  
BEC occurs only if $\bar{d}>2$. An example of this behaviour is given by the  
ladder graph (figure 8) which can be obtained from the  
linear chain by the addition of finite-range links.  
On this structure then the pure hopping model does not exhibit BEC.  
 
An important properties of $\bar{d}$ is that the dimension of the  
graph obtained as a direct product of two graphs is the sum of the  
dimensions of the original structures. An  
example is illustrated in figure 9,  
where we show the direct product of a linear chain and a well known  
fractal, a Sierpinski gasket. In this the coordination number 
case $z_i$ is constant, and applying the previous properties it is easy to show  
that $\bar{d}=1+2\ln(3)/\ln(4)>2$, and therefore,  
applying the previous results, one immediately infers that BEC occurs on this graph.  
  
\section{Concluding remarks} 
We studied in detail BEC of non-interacting bosons on general 
discrete networks, i. e. on graphs.  
We evidenced that these systems support spatial condensation of bosons  
into a single state even if ${\bar d}<2$.  
 
Due to recent advances in engineering technology,  
Josephson junction networks can be build in a variety of controllable,  
non-conventional  
architectures. The theoretical interest for these systems stems  
naturally from the ability of experimentalists in varying the properties  
of the networks by acting directly on their geometry. We expect that  
non universal quantities - such as the critical temperature for  
condensation - will depend crucially on the geometry of a graph,  
thus leading to the possibility of engineering network geometries  
suitable to observe spatial BEC of Cooper pairs. 
 
Our analysis shows that classical inhomogeneous Josephson junction  
networks build on graphs may support a spatial BEC of Cooper pairs induced  
only by the geometry of the ambient space on  
which the junctions lie. We feel our prediction of the existence of spatial BEC  
in  Josephson junction networks may be brought soon to  
direct experimental testing.  
 
{\bf Acknowledgements} 
We thank F. Illuminati and A. Trombettoni for enlightening discussions and  
a careful reading of the manuscript. We thank F. Illuminati for suggesting  
the star graph as an interesting geometrical structure.

\begin{figure}
\begin{center}
\epsfig{file=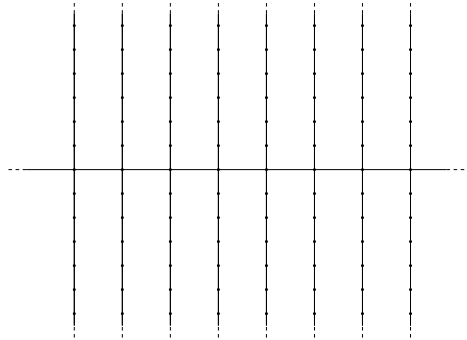, height=4.5cm}
\caption{The comb lattice.}
\end{center}
\end{figure}

\begin{figure}
\begin{center}
\epsfig{file=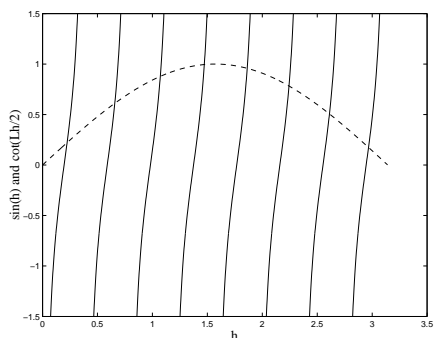, height=4.5cm}
\caption{The numerical solution of equation (\ref{eq7b}) for $k_0=0$ and
$L=16$. We represent the function $\cot(Lh/2)$ with a continuous line and
$\sin(h)$ with a dashed line. The intersections of the two curves are the $L/2$
solutions of (\ref{eq7b}).}
\end{center}
\end{figure}

\begin{figure}
\begin{center}
\epsfig{file=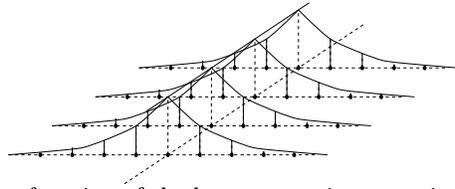, height=6cm, angle=-90}
\caption{The wave function of the low energy eigenstate in the comb lattice.}
\end{center}
\end{figure}

\begin{figure}
\begin{center}
\epsfig{file=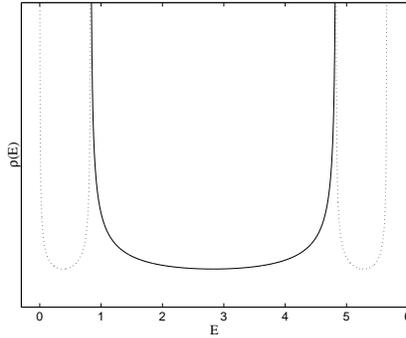, height=4.5cm}
\caption{The spectrum of the pure hopping model on the comb graph. $\rho(E)$
(normalized to $L^2$) is plotted with a continuous line, while for the hidden
spectra (normalized to $L$) we used the dashed line.}
\end{center}
\end{figure}

\begin{figure}
\begin{center}
\epsfig{file=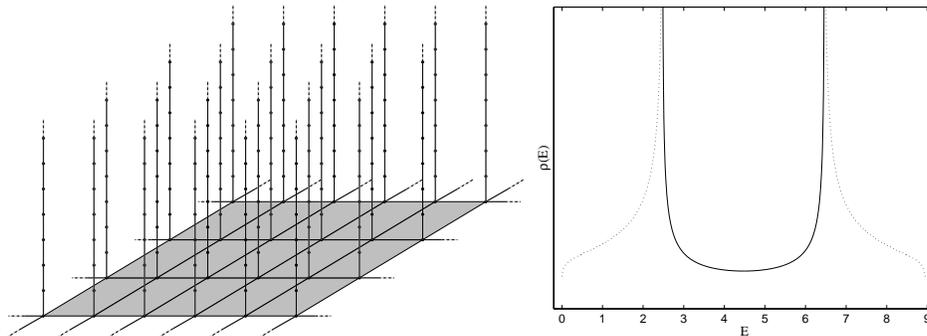, height=4.5cm}
\caption{The brush graph and the spectrum of the pure hopping model.
As in the comb graph, different normalizations are used for the hidden spectra
(dotted line) and $\rho(E)$ (continuous line).}
\end{center}
\end{figure}

\begin{figure}
\begin{center}
\epsfig{file=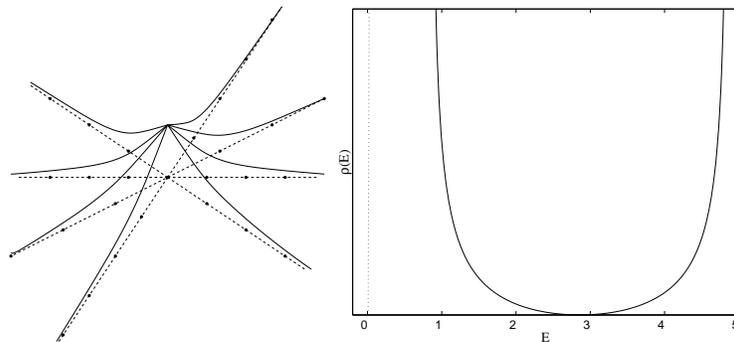, height=4.5cm}
\caption{The star graph with the wave function of the lowest energy state
and the spectrum of the pure hopping model on this structure.}
\end{center}
\end{figure}

\begin{figure}
\begin{center}
\epsfig{file=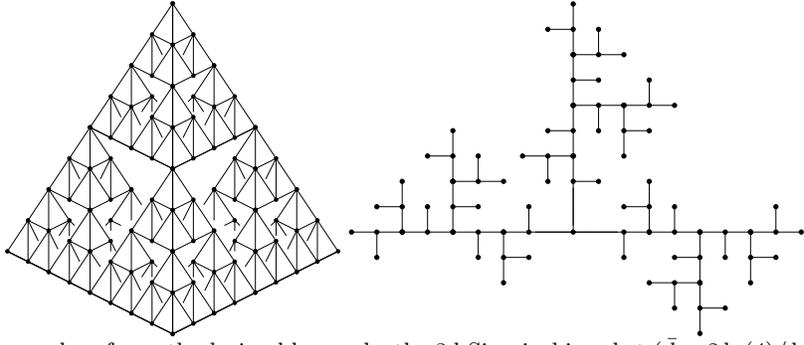, height=4.5cm}
\caption{Two classical examples of exactly decimable graph: the 3d Sierpinski
gasket ($\bar{d}=2\ln(4)/\ln(6)<2$) and the t-fractal
($\bar{d}=2\ln(3)/\ln(6)<2$).}
\end{center}
\end{figure}

\begin{figure}
\begin{center}
\epsfig{file=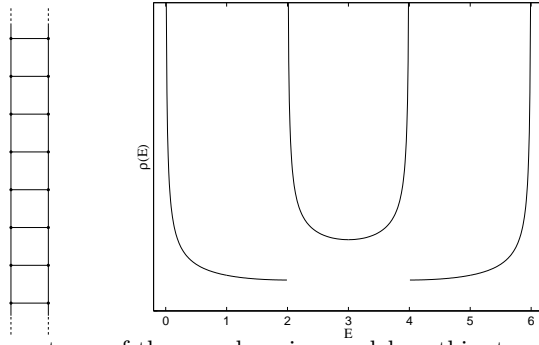, height=4.5cm}
\caption{The ladder graph and the spectrum of the pure hopping model on this
structure. Here we do not have any hidden spectrum and $\rho(E)\to\infty$ when
$E\to E_m=E_0$ ($\bar{d}=1$).}
\end{center}
\end{figure}

\begin{figure}
\begin{center}
\epsfig{file=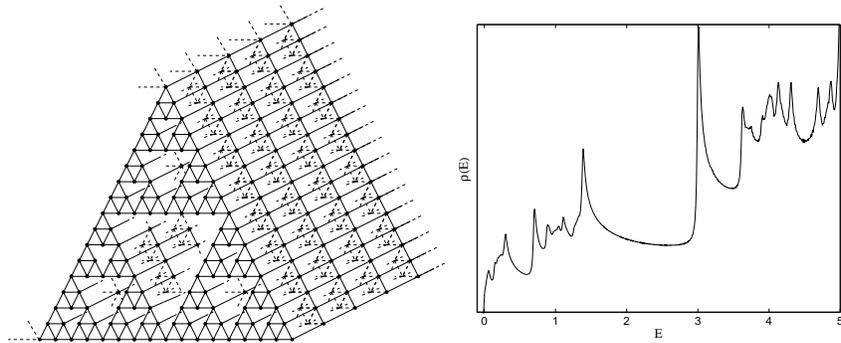, height=4.5cm}
\caption{The graph obtained as a product of the Sierpinski gasket and a linear
chain. Since the coordination number is constant, there is not
hidden spectra in the low energy region. However $\bar{d}=1+2\ln(3)/\ln(4)>2$
(sum of the dimension of the original graphs) and therefore
BEC occurs on this structure.}
\end{center}
\end{figure}
 
\end{document}